\documentclass{elsart}
\usepackage{graphicx}

\newcommand{\order}[1]{\mathcal{O}\left( #1 \right)}
\newcommand{\ep}{\varepsilon}

\begin{document}

\begin{frontmatter}
\begin{flushleft}
ALBERTA-THY-06-07\\
TTP07-14\\
SFB/CPP-07-35
\end{flushleft}
\title{Three-Loop Chromomagnetic Interaction in HQET}
\author[Edmonton,Novosibirsk]{A.G.~Grozin},
\author[Karlsruhe]{P.~Marquard},
\author[Edmonton,Karlsruhe]{J.H.~Piclum} and
\author[Karlsruhe]{M.~Steinhauser}
\address[Edmonton]{Department of Physics, University of Alberta,
Edmonton, Alberta T6G 2G7, Canada}
\address[Novosibirsk]{Budker Institute of Nuclear Physics,
Novosibirsk 630090, Russia}
\address[Karlsruhe]{Institut f\"ur Theoretische Teilchenphysik,
Universit\"at Karlsruhe, 76128 Karlsruhe, Germany}
\begin{abstract}
  We compute the three-loop QCD corrections to the quark chromomagnetic
  moment and thus obtain the matching coefficient and the anomalous
  dimension of the chromomagnetic interaction in HQET. As a byproduct we
  obtain the three-loop corrections to the quark anomalous magnetic
  moment.
\end{abstract}
\begin{keyword}
Heavy Quark Effective Theory \sep radiative corrections
\PACS 12.39.Hg \sep 12.38.Bx
\end{keyword}
\end{frontmatter}


\section{Introduction}
\label{sec::intro}

We consider Quantum Chromodynamics (QCD) with $n_l$ light flavours and
one heavy flavour $Q$. The interaction of a single heavy quark having
momentum $m_Q v+k$ ($m_Q$ is the on-shell mass and $v^2 = 1$) with
gluons and light quarks in the situation when the residual momentum
$k\ll m_Q$ (and momenta of light fields are also small) is described by
the Heavy Quark Effective Theory (HQET)
Lagrangian~\cite{Eichten:1990vp,Falk:1990pz}
\begin{equation}
L = \bar{Q}_v i v\cdot D Q_v
+ \frac{1}{2 m_Q} \left( O_k + C_{cm}(\mu) O_{cm}(\mu) \right)
+ \order{\frac{1}{m_Q^2}}\,,
\label{HQETLagr}
\end{equation}
where $\FMslash{v} Q_v=Q_v$ is the HQET quark field and $D$ denotes the
covariant derivative (see the books~\cite{Manohar:2000dt,Grozin:2004yc}
for more details). The kinetic energy operator
\begin{equation}
O_k = - \bar{Q}_v D_\bot^2 Q_v\,,\qquad
D_\bot = D - v (v\cdot D)
\,,
\label{Ok}
\end{equation}
does not renormalize and its coefficient is equal to 1,
to all orders, due to reparametrization invariance~\cite{Luke:1992cs}.
The chromomagnetic interaction operator is defined as
\begin{equation}
O_{cm} = \frac{1}{2} \bar{Q}_v G_{\mu\nu} \sigma^{\mu\nu} Q_v\,,
\label{Om}
\end{equation}
where $G_{\mu\nu}= g_s G^a_{\mu\nu} t_a$ is the gluon field strength
tensor, $t_a$ is a colour matrix, $\alpha_s=g_s^2/(4\pi)$ the strong coupling
and $\sigma^{\mu\nu} = \frac{i}{2}[\gamma^\mu,\gamma^\nu]$. It is
responsible for the violation of the heavy-quark spin symmetry and thus,
e.g., for the $B$--$B^*$ mass splitting.
The coefficient $C_{cm}(\mu)$ is obtained by matching the scattering
amplitudes of an on-shell heavy quark in an external chromomagnetic field,
expanded in the momentum transfer $q$ up to the linear term,
in the full theory (QCD) and the effective theory (HQET).
If all flavours except $Q$ are massless,
all loop corrections in HQET vanish.
It is most convenient to calculate the QCD scattering amplitude
using the background field method~\cite{Abbott:1980hw}.

The chromomagnetic interaction coefficient $C_{cm}(\mu)$ has been calculated
at one-loop order in Ref.~\cite{Eichten:1990vp}.
The one-loop anomalous dimension $\gamma_{cm}$
of the chromomagnetic operator~(\ref{Om}) follows from its $\mu$ dependence;
it has also been found in Ref.~\cite{Falk:1990pz}.
The two-loop anomalous dimension
has been obtained in Refs.~\cite{Amoros:1997rx,Czarnecki:1997dz},
and the two-loop matching coefficient $C_{cm}(\mu)$ in Ref.~\cite{Czarnecki:1997dz}.
All orders of perturbation theory in the large-$\beta_0$ limit
were summed in Ref.~\cite{Grozin:1997ih}.
The effect of a non-zero charm quark mass 
$m_c$ on the bottom-quark chromomagnetic interaction
at two loops has been investigated in Ref.~\cite{Davydychev:1998si}.
In this paper (cf. Section~\ref{sec::chromo}) we calculate $\gamma_{cm}$,
as well as $C_{cm}(\mu)$ at three loops,
provided that all light flavours are massless.
Using these results,
we obtain the next-to-next-to-leading perturbative correction
to the ratio
\begin{equation}
R = \frac{m_{B^*}^2-m_B^2}{m_{D^*}^2-m_D^2}
\,,
\label{ratio}
\end{equation}
which we discuss in Section~\ref{sec::app}.

The mass difference between the vector and pseudo-scalar $B$ meson is
also often studied with lattice gauge theory simulations
where non-perturbative results
for the operator matrix element can be obtained
(see, e.g., Ref.~\cite{Sommer:2006sj} for an introduction to lattice
HQET and Ref.~\cite{Guazzini:2007bu} for a recent study).
In principle there are two possibilities to
make contact with the experimentally measured
result: the matching can be performed perturbatively and
non-pertubatively. In the first case
the perturbatively computed $n$-loop matching coefficient 
and the $(n+1)$-loop result for the corresponding anomalous dimension 
have to be combined with the non-perturbative lattice results. 
Currently this is done for $n=1$ which
according to Ref.~\cite{Guazzini:2007bu} 
induces an uncertainty of about 4\%. With
the new results of this paper this uncertainty can be significantly
reduced. This topic is dicussed in Section~\ref{sec::lattice}.

We also investigate the heavy-quark magnetic moments.
They have not yet been measured experimentally, however,
for the bottom and the lighter quarks there are upper
limits from LEP1 data~\cite{Escribano:1993xr}.
For the bottom quark this limit is close
to the Standard Model (SM) prediction including the two-loop 
QCD correction~\cite{Bernreuther:2005gq}.
Thus, a more precise measurement at a
future linear collider should be able to determine the bottom-quark
anomalous magnetic moment and probe possible deviations from the 
SM.

The top-quark magnetic moment has not been measured so far. However,
such a measurement would be very interesting since the top-quark
couplings to photons or Z bosons are very sensitive to contributions
from physics beyond the SM. Having this in mind, it is
mandatory to have precise SM predictions for these
couplings.
In Section~\ref{sec::mag} we provide the three-loop QCD corrections to
the coupling of the photon to heavy quarks.


\section{The Calculation}
\label{sec::calc}

\begin{figure}[tbp]
\includegraphics[width=\textwidth]{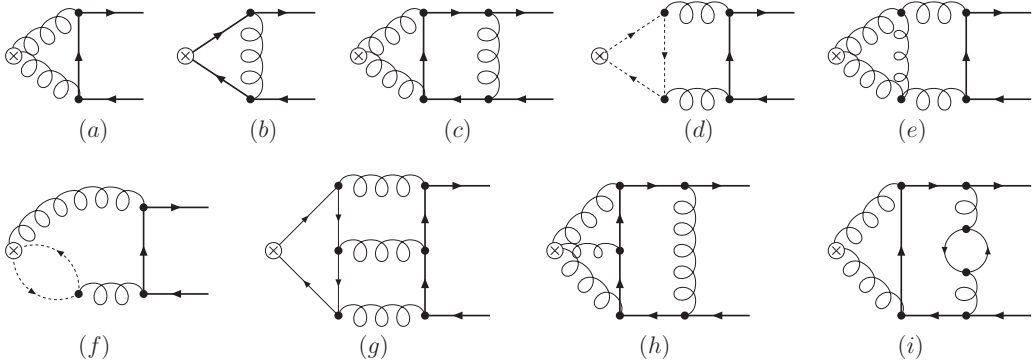}
\caption{Sample diagrams contributing to the
  quark chromomagnetic moment. Solid, curly and dotted lines denote
  quarks, gluons and ghosts, respectively. $\otimes$~represents the
  coupling of the background field. In the closed quark loops all
  flavours have to be considered.}
\label{fig::dias}
\end{figure}

To calculate the chromomagnetic moment we have to consider
the quark--anti-quark--gluon vertex in the background-field formalism in QCD.
Sample diagrams are depicted in Fig.~\ref{fig::dias}.
When both the quark and anti-quark are on the (renormalized) mass shell
and have physical polarizations, the vertex
$\Gamma^\mu_a = \Gamma^\mu t_a$ can be decomposed
into two form factors,
\begin{equation}
  \Gamma^\mu = \gamma^\mu\, F_1(q^2)
    - \frac{i}{2 m_Q} \sigma^{\mu\nu} q_\nu F_2(q^2) \,,
  \label{eq::tensor}
\end{equation}
where $q = p_1 - p_2$ is the gluon momentum and $p_1$ and $p_2$ are the
momenta of the quark and anti-quark, respectively.

The anomalous chromomagnetic moment is given by $\mu_c = Z_2^{\rm OS}
F_2(0)$, where $Z_2^{\rm OS}$ is the quark wave function renormalization
constant in the on-shell scheme.
The total quark colour charge is
given by $Z_2^{\rm OS} F_1(0) = 1$. Thus, $F_1(0)$ is
the inverse of the on-shell wave function renormalization constant,
which has been calculated to three-loops in Ref.~\cite{Melnikov:2000zc}
(see also Ref.~\cite{Marquard:2007uj}). Therefore, the calculation of
$F_1(0)$ provides a strong check on the correctness of our result.

In order to extract the form factors, we use projection operators.
They are conveniently obtained by introducing the momentum
$p=(p_1+p_2)/2$, since $p\cdot q = 0$. With this definition we have
\begin{eqnarray}
  F_1(q^2) &=& \frac{1}{2(d-2)(q^2-4m_Q^2)} \nonumber\\
  && \times \mathop{\mathrm{Tr}}
  \left\{ \left( \FMslash{p}_1 + m_Q \right) \left(
  \gamma_\mu + \frac{4m_Q(d-1)}{q^2-4m_Q^2}\, p_\mu \right) \left(
  \FMslash{p}_2 + m_Q \right)\, \Gamma^\mu \right\}\,,
  \label{eq::proj_F1} \\
  F_2(q^2) &=& -\frac{2m_Q^2}{(d-2)q^2(q^2-4m_Q^2)} \nonumber\\
  && \times \mathop{\mathrm{Tr}}
  \left\{ \left( \FMslash{p}_1 + m_Q \right) \left(
  \gamma_\mu + \frac{4m_Q^2 + (d-2)q^2}{m_Q(q^2-4m_Q^2)}\, p_\mu \right)
  \left( \FMslash{p}_2 + m_Q \right)\, \Gamma^\mu \right\}.
  \label{eq::proj_F2}
\end{eqnarray}
Since the projector for $F_2$ develops a pole for $q^2=0$, we cannot set
$q^2=0$ from the beginning. Instead, we expand in $q$ and keep all terms
which are at most quadratic in $q$. In the final result the limit
$q^2=0$ can be taken. Due to the expansion in $q$ all occurring
integrals are on-shell propagator-type integrals.

All Feynman diagrams are generated with {\tt
QGRAF}~\cite{Nogueira:1991ex} and the various topologies are identified
with the help of {\tt q2e} and {\tt
exp}~\cite{Harlander:1997zb,Seidensticker:1999bb}.
In a next step the reduction of the various functions to so-called master
integrals has to be achieved. For this step we use the so-called
Laporta method~\cite{Laporta:1996mq,Laporta:2001dd}
which reduces the three-loop integrals to 19
master integrals. We use the implementation of Laporta's
algorithm in the program {\tt Crusher}~\cite{PMDS}.
It is written in {\tt C++} and uses
{\tt GiNaC}~\cite{Bauer:2000cp} for simple
manipulations like taking derivatives of polynomial quantities. In the
practical implementation of the Laporta algorithm one of the most
time-consuming operations is the simplification of the coefficients
appearing in front of the individual integrals. This task is performed
with the help of {\tt Fermat}~\cite{fermat} where a special interface
has been used (see Ref.~\cite{Tentyukov:2006ys}).
The main features of the
implementation are the automated generation of the
integration-by-parts (IBP) identities~\cite{Chetyrkin:1981qh}
and a complete symmetrization of the diagrams.
The master integrals are known from~\cite{Melnikov:2000zc} (see also
comments in Ref.~\cite{Marquard:2007uj}).
To calculate the colour factors, we have used the program described in
Ref.~\cite{vanRitbergen:1998pn}.\footnote{We thank Philipp Kant for
  providing his interface for the program of
  Ref.~\cite{vanRitbergen:1998pn}.}

The calculation is performed for an arbitrary gauge parameter in order
to use its cancellation as a check. However, at three-loop level
the expressions for the individual diagrams become very big. Thus
we discard all terms with more than linear $\xi$ dependence. This also
concerns the factors $1/(1-\xi)$ appearing in the vertex of a
background field with two quantum gluons, which does not appear in the
usual formulation of QCD. If this is done, our
final result is gauge-parameter independent up to terms which are
quadratic in $\xi$. Furthermore, our calculation of $F_1$ reproduces
$Z_2^{\rm OS}$ --- including the gauge-dependent
terms~\cite{Melnikov:2000zc,Marquard:2007uj}.


\section{Chromomagnetic moment}
\label{sec::chromo}

The renormalized scattering amplitude of an on-shell heavy quark
with initial momentum $p_1 = m_Q v$
and final momentum $p_2 = m_Q v - q$
in an external gluon field is given by the vertex~(\ref{eq::tensor})
sandwiched between $\bar{u}(p_2)$ and $u(p_1)$
and multiplied by $Z_2^{\rm OS}$.
We expand this amplitude in $q$ up to linear terms,
and re-express (relativistic) QCD spinors
via HQET (non-relativistic) spinors:
\begin{equation}
u(m_Q v+k) = \left[1 + \frac{\rlap/k}{2m_Q}
+ \order{\frac{k^2}{m_Q^2}} \right] u_v(k)\,.
\end{equation}
Then the QCD scattering amplitude reads
\begin{equation}
\bar{u}_v(-q) \left[ v^\mu - \frac{q^\mu}{2m_Q}
- \frac{i}{2m_Q} \sigma^{\mu\nu} q_\nu (1+\mu_c) \right] t_a u_v(0)
\,,
\label{SQCD}
\end{equation}
(we have used $\bar{u}(p_2) \rlap/q u(p_1)=0$).
It must be reproduced by the HQET Lagrangian of Eq.~(\ref{HQETLagr}).
If all flavours except $Q$ are massless,
all loop corrections 
vanish,\footnote{We imply the use of Dimensional Regularization.}
and the scattering amplitude is given by the Born approximation:
\begin{equation}
\bar{u}_v(-q) \left[ v^\mu - \frac{q^\mu}{2m_Q}
- \frac{i}{2m_Q} \sigma^{\mu\nu} q_\nu C_{cm}^0 \right] t_a u_v(0)\,,
\label{SHQET}
\end{equation}
where $C_{cm}^0 = Z_{cm}^{-1}(\mu) C_{cm}(\mu)$
is the bare chromomagnetic interaction coefficient,
and $Z_{cm}(\mu)$ is the $\overline{\mbox{MS}}$ renormalization constant
of the chromomagnetic operator given in Eq.~(\ref{Om}).
Note that both scattering amplitudes~(\ref{SQCD}) and~(\ref{SHQET})
are renormalized and hence ultraviolet-finite, however,
both have infrared divergences.
These divergences are the same,
because HQET has been constructed to reproduce
the infrared behaviour of QCD.
Vanishing loop correction in HQET have ultraviolet and infrared
divergences which cancel each other.
The ultraviolet divergences of $C^0_{cm}$ are removed by $Z_{cm}^{-1}(\mu)$;
the infrared ones match those of $1+\mu_c$ (cf. Eq.~(\ref{SQCD})).

In order to find the chromomagnetic interaction coefficient $C_{cm}(\mu)$
in the HQET Lagrangian~(\ref{HQETLagr}),
we calculate the anomalous chromomagnetic moment
$\mu_c = Z_2^{\rm OS} F_2(0)$
and re-express it in terms of $\alpha_s^{(n_l)}(\mu)$, where the
superscript denotes the number of active flavours. Then
\begin{equation}
  C_{cm}(\mu) = Z_{cm}(\alpha_s^{(n_l)}(\mu))\,
  \left[1 + \mu_c(\alpha_s^{(n_l)}(\mu))\right]\,,
  \label{eq::cm}
\end{equation}
where $n_l = n_f - 1$
is the number of light-quark flavours, which are considered to
be massless in our calculation, and $n_f$ is the total number of 
quark flavours.

The coupling constants
$\alpha_s^{(n_l+1)}(\mu)$ in QCD (with $n_l+1$ flavours)
and $\alpha_s^{(n_l)}(\mu)$ in HQET (with $n_l$ flavours)
are related by~\cite{Chetyrkin:1997un}
\begin{eqnarray}
  &&\hspace{-6mm}
  \frac{\alpha_s^{(n_l+1)}(\mu)}{\pi} = \frac{\alpha_s^{(n_l)}(\mu)}{\pi}
  + \left( \frac{\alpha_s^{(n_l)}(\mu)}{\pi} \right)^2 T_F \bigg[ \frac{1}{3} L +
  \left( \frac{1}{6} L^2 + \frac{1}{36} \pi^2 \right) \ep \nonumber\\
  &&\hspace{-6mm}\qquad{}
  + \left( \frac{1}{18} L^3 + \frac{1}{36} \pi^2 L
  - \frac{1}{9} \zeta_3 \right) \ep^2 + \order{\ep^3} \bigg]
  \nonumber\\
  &&\hspace{-6mm}
  + \left( \frac{\alpha_s^{(n_l)}(\mu)}{\pi} \right)^3 T_F \bigg\{
  \left( \frac{1}{4} L +\frac{15}{16} \right) C_F
  + \left( \frac{5}{12} L - \frac{2}{9} \right) C_A
  + \frac{1}{9} T_F L^2
  \nonumber\\
  &&\hspace{-6mm}
  + \bigg[ \left( \frac{1}{4} L^2 + \frac{15}{8} L
  + \frac{1}{48} \pi^2 + \frac{31}{32} \right) C_F
  + \left( \frac{5}{12} L^2 - \frac{4}{9} L
  + \frac{5}{144} \pi^2 + \frac{43}{108} \right) C_A
  \nonumber\\
  &&\hspace{-6mm}\qquad{}
  + \left( \frac{1}{9} L^3 + \frac{1}{54} \pi^2 L \right) T_F
  \bigg] \ep + \order{\ep^2} \bigg\} + \order{\alpha_s^4}\,,
  \label{Decoupling}
\end{eqnarray}
where $L =\ln(\mu^2/m_Q^2)$ and $m_Q$ is the on-shell mass of the heavy
quark. $C_F= (N_c^2 - 1)/(2N_c)$ and $C_A= N_c$ 
are the eigenvalues of the quadratic Casimir operators
of the fundamental and adjoint representation
for the SU$(N_c)$ colour group, respectively.
In the case of QCD we have $N_c=3$ and $T_F = 1/2$.
$\zeta_n$ denotes Riemann's zeta
function with integer argument $n$.

The ultraviolet divergences contained in $Z_{cm}$ can be transformed
into an anomalous dimension which is given by
\begin{eqnarray}
  \gamma_{cm} &=& \frac{\d\ln Z_{cm}}{\d\ln \mu}
  = \frac{\alpha_s^{(n_l)}}{\pi} \frac{1}{2} C_A + \left(
  \frac{\alpha_s^{(n_l)}}{\pi} \right)^2 C_A \left( \frac{17}{36} C_A -
  \frac{13}{36} T_F n_l \right) \nonumber\\
  &&{} + \left( \frac{\alpha_s^{(n_l)}}{\pi}\right)^3 \bigg\{ \left(
  \frac{1}{8} \zeta_3 + \frac{899}{1728} \right) C_A^3 + \frac{1}{2}
  \pi^2\, \frac{d_F^{abcd}d_A^{abcd}}{C_F N_F} \nonumber\\ 
  &&{} - \bigg[ \left( \frac{1}{2} \zeta_3 + \frac{65}{216} \right) C_A^2
  - \left( \frac{1}{2} \zeta_3 - \frac{49}{96} \right) C_A C_F +
  \frac{1}{36} C_A T_F n_l \bigg] T_F n_l \nonumber\\
  &&{} - \frac{2}{3} \pi^2\, \frac{d_F^{abcd}d_F^{abcd}}{C_F N_F} n_l \bigg\}
  + \order{\alpha_s^4}\,,
  \label{eq::gamma}
\end{eqnarray}
where $d_F^{abcd}$ and $d_A^{abcd}$ are the symmetrized traces
of four generators in the fundamental and adjoint representation, respectively
(for SU$(N_c)$,
$d_F^{abcd}d_F^{abcd}=(N_c^2-1)(N_c^4-6N_c^2+18)/(96N_c^2)$,
$d_F^{abcd}d_A^{abcd}=N_c(N_c^2-1)(N_c^2+6)/48$).
$N_F=N_c$ is the dimension of the fundamental representation.
The two-loop result agrees with~\cite{Amoros:1997rx,Czarnecki:1997dz},
and the $n_l^2$ part of the three-loop one with~\cite{Grozin:1997ih}.

Our result for $C_{cm}$ reads
\begin{eqnarray}
  C_{cm}(\mu) &=& 1 + \frac{\alpha_s^{(n_l)}(m_Q)}{\pi} \left[ \left(
  \frac{1}{4} L + \frac{1}{2} \right) C_A + \frac{1}{2} C_F \right]
  \nonumber\\
  &&{} + \left( \frac{\alpha_s^{(n_l)}(m_Q)}{\pi} \right)^2 \bigg[
  \left( - \frac{1}{2} \pi^2 \ln 2 + \frac{3}{4} \zeta_3
    + \frac{5}{12} \pi^2 - \frac{31}{16} \right) C_F^2 \nonumber\\
  &&{} + \left( \frac{1}{8} L + \frac{1}{12} \pi^2 \ln 2 - \frac{1}{8} \zeta_3
    + \frac{1}{12} \pi^2 + \frac{269}{144} \right) C_F C_A \nonumber\\
  &&{} + \left( - \frac{1}{12} L^2 + \frac{13}{36} L
    + \frac{1}{12} \pi^2 \ln 2 - \frac{1}{8} \zeta_3
    - \frac{17}{144} \pi^2 + \frac{805}{432} \right) C_A^2 \nonumber\\
  &&{} - \frac{25}{36} C_F T_F n_l
    + \left( \frac{1}{24} L^2 - \frac{13}{72} L
    - \frac{1}{36} \pi^2 - \frac{299}{432} \right) C_A T_F n_l \nonumber\\
  &&{} + \left( - \frac{1}{3} \pi^2 + \frac{119}{36} \right) C_F T_F
  + \left( \frac{5}{72} \pi^2 - \frac{149}{216} \right) C_A T_F \bigg] \nonumber\\
  &&{} + \left( \frac{\alpha_s^{(n_l)}(m_Q)}{\pi} \right)^3\, c_{cm}^{(3)}
  + \order{\alpha_s^4}\,.
  \label{eq::Cm}
\end{eqnarray}
The two-loop corrections were already calculated in
Ref.~\cite{Czarnecki:1997dz}.\footnote{Note that in
  Ref.~\cite{Czarnecki:1997dz} the term $\pi^2 C_A T_F$ is not
  correct, since the $\order{\ep}$ term 
  in the one-loop decoupling relation for
  $\alpha_s$~(\ref{Decoupling}) has not been taken into account.}
It is convenient to decompose the three-loop contribution in terms of
the different colour structures as
\begin{eqnarray}
  c_{cm}^{(3)} &=&
  X_{FFF} C_F^3 + X_{FFA} C_F^2 C_A + X_{FAA} C_F C_A^2 + X_{AAA} C_A^3
  \nonumber\\
  &&{} + X_{dd}^{FA} \frac{d_F^{abcd}d_A^{abcd}}{C_F N_F}
  + \left( X_{FFl} C_F^2 + X_{FAl} C_F C_A + X_{AAl} C_A^2 \right) T_F n_l
  \nonumber\\
  &&{} + \left( X_{Fll} C_F + X_{All} C_A \right) T_F^2 n_l^2
  + \left( X_{Flh} C_F + X_{Alh} C_A \right) T_F^2 n_l
  \nonumber\\
  &&{} + \left( X_{FFh} C_F^2 + X_{FAh} C_F C_A + X_{AAh} C_A^2 \right) T_F
  \nonumber\\
  &&{} + \left( X_{Fhh} C_F + X_{Ahh} C_A \right) T_F^2
  + \left( X_{\rm sin}^l n_l + X_{\rm sin}^h \right)
  \frac{d_F^{abcd}d_F^{abcd}}{C_F N_F}\,.
  \label{eq::cm3l}
\end{eqnarray}
Our results for the individual terms read
\begin{eqnarray}
X_{FFF} &=& \frac{20}{3} a_4
  + \frac{5}{18} \ln^4 2
  - \frac{5}{18} \pi^2 \ln^2 2
  - \frac{22}{3} \pi^2 \ln 2
  - \frac{235}{24} \zeta_5
  + \frac{103}{72} \pi^2 \zeta_3 \nonumber\\
  &&{} - \frac{139}{2160} \pi^4
  + \frac{241}{24} \zeta_3
  + \frac{23}{6} \pi^2
  - \frac{101}{64}\,,\\
X_{FFA} &=& \bigg( - \frac{1}{8} \pi^2 \ln 2
    + \frac{3}{16} \zeta_3
    + \frac{5}{48} \pi^2
    - \frac{31}{64} \bigg) L
  - \frac{35}{3} a_4
  - \frac{35}{72} \ln^4 2 \nonumber\\
  &&{} + \frac{13}{9} \pi^2 \ln^2 2
  - \frac{101}{24} \pi^2 \ln 2
  + \frac{115}{6} \zeta_5
  - \frac{17}{9} \pi^2 \zeta_3
  - \frac{209}{2880} \pi^4
  - \frac{847}{96} \zeta_3 \nonumber\\
  &&{} + \frac{9767}{1728} \pi^2
  - \frac{2803}{288}\,,\\
X_{FAA} &=& - \frac{1}{24} L^2
  + \bigg( \frac{1}{48} \pi^2 \ln 2
    - \frac{1}{32} \zeta_3
    + \frac{1}{48} \pi^2
    + \frac{337}{576} \bigg) L
  + \frac{191}{18} a_4 \nonumber\\
  &&{} + \frac{191}{432} \ln^4 2
  - \frac{169}{216} \pi^2 \ln^2 2
  + \frac{1745}{432} \pi^2 \ln 2
  - \frac{265}{16} \zeta_5
  + \frac{161}{72} \pi^2 \zeta_3 \nonumber\\
  &&{} + \frac{491}{10368} \pi^4
  + \frac{2951}{288} \zeta_3
  - \frac{17375}{3456} \pi^2
  + \frac{122971}{10368}\,,\\
X_{AAA} &=& \frac{19}{432} L^3
  - \frac{497}{1728} L^2
  + \left( \frac{1}{48} \pi^2 \ln 2
    + \frac{1}{32} \zeta_3
    - \frac{17}{576} \pi^2
    + \frac{2917}{3456} \right) L \nonumber\\
  &&{} - \frac{8}{3} a_4
  - \frac{1}{9} \ln^4 2
  - \frac{1}{36} \pi^2 \ln^2 2
  - \frac{317}{864} \pi^2 \ln 2
  + \frac{925}{192} \zeta_5
  - \frac{653}{864} \pi^2 \zeta_3 \nonumber\\
  &&{} - \frac{17}{810} \pi^4
  - \frac{6079}{1728} \zeta_3
  + \frac{1585}{1296} \pi^2
  + \frac{1302797}{186624}\,,\\
X_{dd}^{FA} &=& \frac{1}{4} \pi^2 L
  - \frac{40}{3} a_4
  - \frac{5}{9} \ln^4 2
  + \frac{20}{9} \pi^2 \ln^2 2
  + \frac{91}{12} \pi^2 \ln 2
  - 10 \zeta_5 \nonumber\\
  &&{} + \frac{97}{72} \pi^2 \zeta_3
  + \frac{7}{270} \pi^4
  - \frac{73}{24} \zeta_3
  - \frac{151}{27} \pi^2
  - \frac{5}{18}\,,\\
X_{FFl} &=& - \frac{8}{3} a_4
  - \frac{1}{9} \ln^4 2
  - \frac{2}{9} \pi^2 \ln^2 2
  + \frac{5}{3} \pi^2 \ln 2
  + \frac{11}{216} \pi^4
  - 3 \zeta_3
  - \frac{79}{54} \pi^2 \nonumber\\
  &&{} + \frac{125}{32}\,,\\
X_{FAl} &=& \frac{5}{96} L^2
  + \biggl( \frac{1}{4} \zeta_3
    - \frac{299}{576} \biggr) L
  + \frac{4}{9} a_4
  + \frac{1}{54} \ln^4 2
  + \frac{1}{27} \pi^2 \ln^2 2 \nonumber\\
  &&{} - \frac{1}{108} \pi^2 \ln 2
  - \frac{23}{3240} \pi^4
  + \frac{2}{3} \zeta_3
  - \frac{23}{72} \pi^2
  - \frac{88351}{10368}\,,\\
X_{AAl} &=& - \frac{35}{864} L^3
  + \frac{235}{864} L^2
  - \left( \frac{1}{4} \zeta_3
    + \frac{1}{144} \pi^2
    + \frac{715}{1728} \right) L
  + \frac{4}{9} a_4
  + \frac{1}{54} \ln^4 2 \nonumber\\
  &&{} + \frac{1}{27} \pi^2 \ln^2 2
  - \frac{89}{216} \pi^2 \ln 2
  + \frac{1}{180} \pi^4
  - \frac{101}{432} \zeta_3
  + \frac{35}{864} \pi^2 \nonumber\\
  &&{} - \frac{236801}{46656}\,,\\
X_{Fll} &=& \frac{1}{27} \pi^2
  + \frac{317}{324}\,,\\
X_{All} &=& \frac{1}{108} L^3
  - \frac{13}{216} L^2
  - \frac{1}{72} L
  + \frac{7}{54} \zeta_3
  + \frac{25}{324} \pi^2
  + \frac{3535}{5832}\,,\\
X_{Flh} &=& \frac{1}{27} \pi^2
  - \frac{61}{162}\,,\\
X_{Alh} &=& - \frac{11}{108} \pi^2
  + \frac{167}{162}\,,\\
X_{FFh} &=& \frac{32}{3} a_4
  + \frac{4}{9} \ln^4 2
  - \frac{4}{9} \pi^2 \ln^2 2
  - \frac{16}{9} \pi^2 \ln 2
  + \frac{4}{135} \pi^4
  - \frac{263}{72} \zeta_3 \nonumber\\
  &&{} + \frac{11}{162} \pi^2
  + \frac{2027}{216}\,,\\
X_{FAh} &=& \left( - \frac{1}{12} \pi^2
    + \frac{119}{144} \right) L
  - 10 a_4
  - \frac{5}{12} \ln^4 2
  + \frac{5}{12} \pi^2 \ln^2 2
  + \frac{83}{27} \pi^2 \ln 2 \nonumber\\
  &&{} - \frac{25}{24} \zeta_5
  + \frac{1}{8} \pi^2 \zeta_3
  - \frac{101}{1440} \pi^4
  - \frac{6937}{864} \zeta_3
  - \frac{22241}{19440} \pi^2
  + \frac{8447}{864}\,,\\
X_{AAh} &=& \left( \frac{5}{288} \pi^2
    - \frac{149}{864} \right) L
  + a_4
  + \frac{1}{24} \ln^4 2
  - \frac{1}{24} \pi^2 \ln^2 2
  + \frac{1211}{432} \pi^2 \ln 2 \nonumber\\
  &&{} - \frac{65}{144} \zeta_5
  + \frac{65}{432} \pi^2 \zeta_3
  + \frac{53}{5184} \pi^4
  + \frac{4423}{1728} \zeta_3
  - \frac{283429}{155520} \pi^2 \nonumber\\
  &&{} - \frac{71965}{10368}\,,\\
X_{Fhh} &=& \frac{8}{3} \zeta_3
  - \frac{4}{135} \pi^2
  - \frac{943}{324}\,,\\
X_{Ahh} &=& - \frac{4}{9} \zeta_3
  + \frac{1}{270} \pi^2
  + \frac{487}{972}\,,\\
X_{\rm sin}^l &=& - \frac{1}{3} \pi^2 L
  + \frac{29}{270} \pi^4
  - 3 \zeta_3
  - \frac{44}{27} \pi^2
  + \frac{2}{3}\,,\\
X_{\rm sin}^h &=& 16 a_4
  + \frac{2}{3} \ln^4 2
  - \frac{2}{3} \pi^2 \ln^2 2
  - 24 \pi^2 \ln 2
  + \frac{5}{6} \zeta_5
  - \frac{5}{18} \pi^2 \zeta_3
  - \frac{41}{540} \pi^4 \nonumber\\
  &&{} - \frac{4}{3} \zeta_3
  + \frac{931}{54} \pi^2
  + \frac{5}{9}\,,
\end{eqnarray}
with $a_4 = \mathop{\mathrm{Li}}\nolimits_4(1/2)$. The $n_l^2$ part agrees
with the result obtained in Ref.~\cite{Grozin:1997ih}.

Substituting numerical values of the constants,
we obtain, for the physical SU(3) colour group,
\begin{eqnarray}
  C_{cm}(m_Q) &=& 1 + 0.6897\, \alpha_s^{(n_l)}(m_Q)
  + \left( 2.2186 - 0.1938\,n_l \right) \left[\alpha_s^{(n_l)}(m_Q)\right]^2
  \nonumber\\
  &&\mbox{} + \left( 11.079 - 1.7490\, n_l + 0.0513\, n_l^2 \right)
  \left[\alpha_s^{(n_l)}(m_Q)\right]^3 + \order{\alpha_s^4}
  \nonumber\\
  &=&{}1 + 0.6897\, \alpha_s^{(n_l)}(m_Q)
  + \left( 1.1626\,\beta_0 - 0.9786 \right)
  \left[\alpha_s^{(n_l)}(m_Q)\right]^2
  \nonumber\\
  &&\mbox{} + \left( 1.8468\,\beta_0^2 + 0.3370\,\beta_0 - 3.8137 \right)
  \left[\alpha_s^{(n_l)}(m_Q)\right]^3 + \order{\alpha_s^4}\,,
  \nonumber\\
\end{eqnarray}
with $\beta_0 = (11 C_A/3 - 4T_F n_l/3)/4$.
The first-order $1/\beta_0$ result~\cite{Grozin:1997ih}
contains the highest powers of $\beta_0$ in each term.
For $n_l=4$, for example, the coefficient
of $\left[\alpha_s^{(n_l)}(m_Q)\right]^2$ is $2.4221-0.9786=1.4435$,
and that of $\left[\alpha_s^{(n_l)}(m_Q)\right]^3$
is $8.0155+0.7020-3.8137=4.9039$;
the large-$\beta_0$ approximation of Ref.~\cite{Grozin:1997ih},
which only includes the first terms in these sums,
overestimates these two coefficients by 68\% and 63\%,
correspondingly.

For the numerical evaluation of $C_{cm}(m_Q)$ and $\gamma_{cm}$, we 
use the values $m_c=1.6$~GeV, $m_b=4.7$~GeV and
$m_t=175$~GeV. The number of light-quark flavours $n_l$ is three, four
and five for the charm, bottom and top quark, respectively.
To evaluate
$\alpha_s^{(n_l)}(m_Q)$, defined with $n_l$ active flavours, from
$\alpha_s^{(5)}(m_Z)=0.118$, we use the
program {\tt RunDec}~\cite{Chetyrkin:2000yt} and obtain
$\alpha_s^{(3)}(m_c)=0.3348$, 
$\alpha_s^{(4)}(m_b)=0.2163$ and $\alpha_s^{(5)}(m_t)=0.1074$,
which leads to
\begin{eqnarray}
  C_{cm}(m_c) &=& 1 + 0.2309 + 0.1835 + 0.2362 = 1.6506\,, \\
  C_{cm}(m_b) &=& 1 + 0.1492 + 0.0676 + 0.0497 = 1.2664\,, \\
  C_{cm}(m_t) &=& 1 + 0.0741 + 0.0144 + 0.0045 = 1.0930\,.
\end{eqnarray}
In the case of the charm quark, we see that the perturbative series does
not converge which is probably connected to the 
relatively light scale of $1.6$~GeV at which the strong coupling is
evaluated. Thus, it seems that there are potentially large
non-perturbative corrections.
While the situation is better in the case of the
bottom quark, the corrections are still very large. The three-loop
correction amounts to about 30\% of the one-loop contribution. For
the top quark, we find that our new term contributes about 6\% of the
one-loop correction leading to a fairly reliable prediction of $C_{cm}(m_t)$.

The numerical evaluation of the anomalous dimension~(\ref{eq::gamma})
gives
\begin{eqnarray}
  \gamma_{cm} &=& 0.4775\, \alpha_s^{(n_l)}
  + \left( 0.4306 - 0.0549\, n_l \right) \left[\alpha_s^{(n_l)}\right]^2
  \nonumber\\
  &&{} + \left( 0.8823 - 0.1472\, n_l - 0.0007\, n_l^2 \right)
  \left[\alpha_s^{(n_l)}\right]^3 + \order{\alpha_s^4}\,.
\end{eqnarray}
For the individual quark flavours this leads to
\begin{eqnarray}
  \gamma_{cm}(m_c) &=& 0.1599 + 0.0298 + 0.0163 = 0.2060\,, 
  \nonumber\\
  \gamma_{cm}(m_b) &=& 0.1033 + 0.0099 + 0.0029 = 0.1160\,,
  \nonumber\\
  \gamma_{cm}(m_t) &=& 0.0513 + 0.0018 + 0.0002 = 0.0533\,.
  \label{eq::gammacm-b}
\end{eqnarray}
Thus, as far as the anomalous dimension is concerned the convergence
behaviour is acceptable even for the charm quark.

These observations are in good agreement
with the analysis of Ref.~\cite{Grozin:1997ih}.
The chromomagnetic interaction coefficient $C_{cm}(m_Q)$
has the leading renormalon singularity (namely, a branch point)
at the Borel parameter $u=1/2$, quite close to the origin;
it leads to a very fast growth of coefficients of the perturbative series,
$\sim n! (\beta_0/2)^n$.
It also means that the leading non-perturbative correction
is only suppressed by the first power of $1/m_Q$,
and is thus important, especially for the charm quark.
On the other hand, the perturbative series for the anomalous dimension
has a finite radius of convergence.


\section{Application: heavy-meson mass splittings}
\label{sec::app}

The most prominent physical effect caused by the chromomagnetic
interaction is the mass splittings of hadronic doublets
which are degenerate at $m_Q=\infty$ due to the heavy-quark
spin symmetry.
For example, for the bottom mesons $B$ and $B^*$ one 
has~\cite{Bigi:1994ga},
\begin{equation}
m_{B^*}^2 - m_B^2 = \frac{4}{3} C_{cm}^{(4)}(\mu) \mu_{G(4)}^2(\mu)
+ \order{\frac{\Lambda_{\rm QCD}}{m_b}}\,,
\label{Splitting}
\label{eq::mBmB}
\end{equation}
where the index ``(4)'' means that we are considering $n_l=4$ flavour
HQET, and $\mu_{G(4)}^2(\mu)$ is the matrix element of $O_{cm}(\mu)$
(cf. Eq.~(\ref{Om})) over the ground-state meson.
It is most natural to choose $\mu=m_b$ in Eq.~(\ref{eq::mBmB}), 
because then $C_{cm}$ contains no large logarithms.
A similar formula can be written down for the $D$ mesons where 
$C_{cm}^{(3)}(\mu)$ and $\mu_{G(3)}^2(\mu)$ appear in the corresponding
expression for $m_{D^*}^2-m_D^2$.
The running of $\mu_{G(n_l)}^2(\mu)$ is governed
by the anomalous dimension given in
Eq.~(\ref{eq::gamma}). Furthermore, it is necessary to relate the 
matrix elements in the two theories via the following
decoupling relation
\begin{eqnarray}
\mu_{G(4)}^2(m_c) &=& \mu_{G(3)}^2(m_c)
\left[ 1 \!+\! z_2 \left(\frac{\alpha_s^{(4)}(m_c)}{\pi}\right)^2 
  \!+\! z_3 \left(\frac{\alpha_s^{(4)}(m_c)}{\pi}\right)^3 
  + \order{\alpha_s^4} \right]\,,
\nonumber\\
\label{CMdecoupling}
\end{eqnarray}
with $z_2=-71C_A T_F/432 $~\cite{Grozin:2000cm,Grozin:2004yc}.
We introduce the unknown\footnote{$z_3$ can be calculated using 3-loop
HQET integrals considered in~\cite{Grozin:2006xm,Grozin:2007ap}.}
coefficient $z_3$ since it appears in 
estimates of higher order effects which are presented below.

In the formulation with resummed logarithms one combines for
consistency Eq.~(\ref{CMdecoupling}) with the two-loop result for $C_{cm}$
and the three-loop anomalous dimension.
This leads to the ratio $R$ of Eq.~(\ref{ratio})
to the next-to-next-to-leading (NNL) order approximation. For
later use we extend the formalism to NNNL order where we obtain
\begin{eqnarray}
  R &=& x^{-\gamma_0/(2\beta_0)} \Biggl\{ 1 + r_1 (x-1)
  \frac{\alpha_s^{(4)}(m_b)}{\pi} \nonumber\\
  &&{} + \left[ r_{20} + r_{21} (x^2-1)
  + \frac{r_1^2}{2} (x-1)^2 \right]
  \left( \frac{\alpha_s^{(4)}(m_b)}{\pi} \right)^2 \nonumber\\
  &&{} + \Biggl[ r_{30} + r_{31} (x-1) (x^2+x+1)
  + \frac{r_1^3}{6} (x-1)^3 + r_1 r_{20} (x-1) \nonumber\\
  &&\hphantom{{}+\Biggl[\biggr.}
  + r_1 r_{21} (x-1)^2 (x+1) \Biggr]
  \left( \frac{\alpha_s^{(4)}(m_b)}{\pi} \right)^3
  + \order{\alpha_s^4,\frac{\Lambda_{\rm QCD}}{m_{b,c}}} \Biggr\}\,,
  \label{eq::RtoNNL}
\end{eqnarray}
with
\begin{eqnarray}
  x &=& \frac{\alpha_s^{(4)}(m_c)}{\alpha_s^{(4)}(m_b)}\,, \nonumber\\
  r_1 &=& - c_{cm}^{(1)} - \frac{\gamma_0}{2\beta_0} \left(
  \frac{\gamma_1}{\gamma_0} - \frac{\beta_1}{\beta_0} \right)\,, \nonumber\\
  r_{20} &=& c_{cm}^{(2)}(n_l=4) - c_{cm}^{(2)}(n_l=3) + z_2\,, \nonumber\\
  r_{21} &=& - c_{cm}^{(2)}(n_l=3) + \frac{\left(c_{cm}^{(1)}\right)^2}{2}
  + z_2 \nonumber\nonumber\\
  &&{} + \frac{\gamma_0}{4\beta_0} \left[ - \frac{\gamma_2}{\gamma_0}
  + \frac{\beta_1}{\beta_0} \frac{\gamma_1}{\gamma_0}
  + \frac{\beta_2}{\beta_0} - \left(\frac{\beta_1}{\beta_0}\right)^2
  \right]\,, \nonumber\\
  r_{30} &=& c_{cm}^{(3)}(n_l=4) - c_{cm}^{(3)}(n_l=3)
  - c_{cm}^{(1)} \left( c_{cm}^{(2)}(n_l=4) - c_{cm}^{(2)}(n_l=3) + d_2 \right)
  \nonumber\\&&\mbox{}
  + z_3
  \,, \nonumber\\
  r_{31} &=& - c_{cm}^{(3)}(n_l=3)
  + c_{cm}^{(1)} \left( c_{cm}^{(2)}(n_l=3) - d_2 \right)
  - \frac{\left(c_{cm}^{(1)}\right)^3}{3}
  + z_3 
  + \frac{\gamma_0}{6\beta_0} \left[- \frac{\gamma_3}{\gamma_0}
    \right.
  \nonumber\\
  &&\left.{} 
  + \frac{\beta_1}{\beta_0} \frac{\gamma_2}{\gamma_0}
  + \frac{\beta_2}{\beta_0} \frac{\gamma_1}{\gamma_0}
  - \left(\frac{\beta_1}{\beta_0}\right)^2 \frac{\gamma_1}{\gamma_0}
  + \frac{\beta_3}{\beta_0}
  - 2 \frac{\beta_1}{\beta_0} \frac{\beta_2}{\beta_0}
  + \left(\frac{\beta_1}{\beta_0}\right)^3 \right]
  \,.
\end{eqnarray}
$c_{cm}^{(n)}$ denotes the coefficient of $(\alpha_s(m_Q)/\pi)^n$
in $C_{cm}(\mu=m_Q)$. 
The terms $z_2$ and $z_3$ stem from the decoupling of the matrix
element and are introduced in Eq.~(\ref{CMdecoupling})
and  $d_2 = [ (2/9) C_A - (15/16) C_F ] T_F$ 
stems from the decoupling of $\alpha_s$
(cf. Eq.~(\ref{Decoupling})).
$\gamma_n$ are the coefficients of $(\alpha_s/\pi)^{n+1}$
in the anomalous dimension and the coefficients of the $\beta$ function
are used in the form $\beta_0 = (11 C_A/3 - 4T_F n_l/3)/4$; see
Refs.~\cite{vanRitbergen:1997va,Czakon:2004bu} for the remaining $\beta_i$.

Both for $\gamma_n$ and $\beta_n$ $n_l=4$ active flavours have to be chosen.
Let us mention that the 
next-to-leading (NL) order result of Eq.~(\ref{eq::RtoNNL})
has been obtained in Ref.~\cite{Amoros:1997rx}.
Inserting the numerical values given above and displaying the
contributions from the individual orders separately, we find
\begin{eqnarray}
  R &=& 0.8517 - 0.0696 - 0.0908 + \left[ -0.1285 \right] + \dots
  \nonumber\\
  &=& 0.6914 + \left[ -0.1285 \right] +
  \dots\,,
  \label{eq::Rsum}
\end{eqnarray}
where the ellipses denote terms of higher order and power
corrections. The term in square brackets is our estimate of the fourth
order contribution, 
where we assume that the four-loop coefficient of the
anomalous dimension is negligible. This is justified by the rapid
convergence of $\gamma_{cm}$ in the case of the bottom quark as can be
seen in Eq.~(\ref{eq::gammacm-b}).
Furthermore, we set the unknown coefficient $z_3$ of
Eq.~(\ref{CMdecoupling}) to zero which is a good approximation since
it enters with a small coefficient. 

The experimental value is $R_{\rm exp} = 0.88$~\cite{pdg} with a 
negligible uncertainty.
The NNL correction amounts to 10\% of the LO contribution, however, it
is larger than the NL one. Furthermore it is negative and thus
increases the difference of the perturbative result and the
experimental value. The estimated third-order correction is even
larger than the NL and NNL one and contributes also with a negative sign.
This indicates
that the $\Lambda_{\rm QCD}/m_c$ correction may be quite substantial.

It is interesting to consider the quantity $R$ also without performing
the resummation of the logarithms. In this way the three-loop result
for the coefficient can be incorporated in a consistent way.
The starting point is Eq.~(\ref{ratio}) where quantities defined for
$n_l=4$ are present in the numerator and the ones defined for $n_l=3$
in the denominator. Using Eq.~(\ref{CMdecoupling}) for the decoupling of
the matrix element and running from $\mu=m_c$ to $\mu=m_b$ cancels 
$\mu_{G(4)}^2$. Afterwards, we can replace $\alpha_s^{(3)}(m_c)$ by
$\alpha_s^{(4)}(m_b)$, using decoupling and renormalization group
running, and perform a consistent expansion of $R$ in
$\alpha_s^{(4)}(m_b)$. As a result we obtain
\begin{eqnarray}
  R &=& 1 - \gamma_0\, l\, \frac{\alpha_s^{(4)}(m_b)}{\pi}
  \nonumber\\
  && + \bigg[
    \gamma_0 \left( \frac{1}{2} \gamma_0 - \beta_0 \right) l^2
    - \left( 2 \beta_0 c_{cm}^{(1)} + \gamma_1 \right) l \nonumber\\
    &&\qquad{} + c_{cm}^{(2)}(n_l=4) - c_{cm}^{(2)}(n_l=3) + z_2
    \bigg] \left( \frac{\alpha_s^{(4)}(m_b)}{\pi} \right)^2
  \nonumber\\
  && + \biggl[
    \gamma_0 \left( - \frac{1}{6} \gamma_0^2 + \beta_0 \gamma_0
    - \frac{4}{3} \beta_0^2 \right) l^3 + \left( (\gamma_0-2\beta_0) ( 2
    \beta_0 c_{cm}^{(1)} + \gamma_1 ) - \beta_1 \gamma_0 \right) l^2
  \nonumber\\
  &&\qquad{} + \bigg( \gamma_0 ( c_{cm}^{(2)}(n_l=3) - c_{cm}^{(2)}(n_l=4) ) -
  4 \beta_0 c_{cm}^{(2)}(n_l=3) + 2 \beta_0 (c_{cm}^{(1)})^2
  \nonumber\\
  &&\qquad{} - 2 \beta_1 c_{cm}^{(1)} + ( 4 \beta_0 - \gamma_0 ) z_2 -
  \gamma_2 \bigg) l + c_{cm}^{(3)}(n_l=4) - c_{cm}^{(3)}(n_l=3)
  \nonumber\\
  &&\qquad{}  - c_{cm}^{(1)} (c_{cm}^{(2)}(n_l=4) - c_{cm}^{(2)}(n_l=3) + d_2 ) +
  z_3 \biggr] \left( \frac{\alpha_s^{(4)}(m_b)}{\pi} \right)^3
  \nonumber\\
  &&{} + \order{\alpha_s^4}\,,
  \label{res}
\end{eqnarray}
where $l = \ln (m_b/m_c)$. In our numerical evaluation we set the
decoupling coefficient $z_3$ to zero.
Since similar three-loop decoupling effects are small we expect the
same in our case. Furthermore, note that $c_{cm}^{(3)}$ is numerically 
rather large. Inserting the numerical values yields
\begin{equation}
  R = 1 - 0.1113 - 0.0780 - 0.0755 + \dots = 0.7352 + \dots
  \,.
  \label{eq::Rnosum}
\end{equation}
Comparing Eqs.~(\ref{eq::Rsum}) and (\ref{eq::Rnosum}), we find that the
convergence of $R$ without resummation behaves slightly better. 
However, the coefficients of the perturbative series are
still large.


\section{Matching coefficient and renormalization group invariant
  quark mass}
\label{sec::lattice}

In this section we would like to discuss the result of the
matching coefficient in the form which is often used in lattice
simulations of the mass difference $m_{B^*}^2-m_B^2$.
In doing so we follow the procedure outlined in
Ref.~\cite{Heitger:2004gb}.

In lattice simulations one usually determines the
renormalization group invariant (RGI) matrix element of
the operator $O_{cm}$ which has to be multiplied by the 
corresponding matching coefficient. For its derivation one considers
in a first step
\begin{equation}
C_{\rm mag} = \left ( 2 \beta_0 \frac{\alpha_s(\bar m_*)}{\pi}
\right)^{\frac{\gamma_0^{\rm mag}}{2\beta_0}} \exp
  \left[
    \int_0^{\alpha_s(\bar m_*)}
    \left ( \frac{\gamma_{\rm mag}}{2 \beta} - \frac{\gamma^{\rm
	mag}_{0}}{2 \beta_0} \right )\frac{d\alpha_s}{\alpha_s}
    \right] ,
\label{eq:1}
  \end{equation}
where $\bar m_* = \bar{m}_Q(\bar m_*)$ is the scale
invariant $\overline{\mbox{MS}}$ mass. The anomalous dimension
$\gamma_{\rm mag}$ is given by a combination of
$\gamma_{cm}(\alpha_s)$, $\beta(\alpha_s)$ and $C_{cm}(\mu)$ and reads 
\begin{eqnarray}
  \gamma_{\rm mag} &=& \frac{\alpha_s}{\pi} \left(\gamma^{\rm mag}_0 +
  \gamma^{\rm mag}_1\, \frac{\alpha_s}{\pi} + \gamma^{\rm mag}_2\,
  \left(\frac{\alpha_s}{\pi}\right)^2  + \ldots\right)
  \\ 
  &=&\gamma_{cm} + 2\beta_0 c_{cm}^{(1)}
  \left(\frac{\alpha_s}{\pi}\right)^2 
  + \left( 4
  \beta_0 c_{cm}^{(2)} + 2 \beta_1 c_{cm}^{(1)} - 2 \beta_0
  (c_{cm}^{(1)})^2\right)
  \left(\frac{\alpha_s}{\pi}\right)^3 
  \nonumber\\&&\mbox{}
  +\left( 2 \beta_2 c_{cm}^{(1)} - 2 \beta_1 (c_{cm}^{(1)})^2 
  + 2 \beta_0 (c_{cm}^{(1)})^3 + 4 \beta_1 c_{cm}^{(2)} 
  \right.\nonumber\\&&\left.\mbox{}
  - 
  6 \beta_0 c_{cm}^{(1)} c_{cm}^{(2)} + 6 \beta_0 c_{cm}^{(3)} \right)  
  \left(\frac{\alpha_s}{\pi}\right)^4
  + \ldots
\nonumber
\end{eqnarray}
The terms containing the $\beta$-function stem from the running of 
$\alpha_s$.

Since $C_{\rm mag}$ is still multiplied by the inverse 
pole mass, $1/m_Q$, in a
further step the renormalization group invariant mass $M^{\rm {RGI}}$
defined by~\cite{Heitger:2004gb}
\begin{eqnarray}
  {M^{RGI}} &=& {\bar m_*}\left( 2\beta_0
  \frac{\alpha_s(\bar m_*)}{\pi}\right)^{-\frac{\gamma_{m,0}}{\beta_0}}
  \exp\left[ - \int_0^{\alpha_s(\bar{m}_*)}
    \left ( \frac{\gamma_{m}}{\beta} - \frac{\gamma_{m,0}}{\beta_0} 
    \right)\frac{d\alpha_s}{\alpha_s}
    \right]
  \label{eq::MRGI}
  \,,
\end{eqnarray}
can be used, where we introduced the quark mass anomalous dimension
$\gamma_m$
\begin{eqnarray}
  \gamma_m &=& \frac{\alpha_s}{\pi} \left(\gamma_{m,0} +
  \gamma_{m,1}\, \frac{\alpha_s}{\pi} + \gamma_{m,2}\,
  \left(\frac{\alpha_s}{\pi}\right)^2  + \ldots\right)
  \,,
\end{eqnarray}
with $\gamma_{m,0} = 3C_F/4$; the other coefficients can be
found in Refs.~\cite{Chetyrkin:1997dh,Vermaseren:1997fq}.
Using Eq.~(\ref{eq::MRGI}) in addition to the $\overline{\rm
  MS}$--on-shell relation the overall and logarithmic dependence on the
pole mass can be replaced by the RGI mass. This procedure turns
$C_{\rm mag}$ into $C_{\rm spin}$ and one obtains an equation
analog to Eq.~(\ref{eq:1}) for $C_{\rm spin}$ with the anomalous
dimension given by
\begin{eqnarray}
  \gamma_{\rm spin} &=& \tilde{\gamma}_{\rm mag} - 2 \gamma_{m}
\end{eqnarray}
with
\begin{eqnarray}
  \tilde{\gamma}^{\rm mag}_0 &=& \gamma^{\rm mag}_0\,, \nonumber\\
  \tilde{\gamma}^{\rm mag}_1 &=& \gamma^{\rm mag}_1 - 2 \beta_0 k_1\,,
  \nonumber\\ 
  \tilde{\gamma}^{\rm mag}_2 &=& \gamma^{\rm mag}_2 - 2 \beta_1 k_1 +
  \beta_0 ( 2 k_1^2 - 4 \gamma_0 k_1 - 4 k_2 )\,, \nonumber\\
  \tilde{\gamma}^{\rm mag}_3 &=& \gamma^{\rm mag}_3 - 2 \beta_2 k_1 +
  \beta_1 ( 2 k_1^2 - 4 \gamma_0 k_1 - 4 k_2) - 12 \beta_0^2 k_1
  c_{cm}^{(1)} \nonumber\\
  && + \beta_0 \left( 3 \gamma_0 k_1^2 - 2k_1^3 - 6 \gamma_1 k_1 + 6 k_1
  k_2 - 6\gamma_0 k_2 - 6 k_3 \right)\,.
\end{eqnarray}
The terms containing the $\beta$-function stem from the transformation
of $m_Q$ to $\bar{m}_Q$ and the coefficients $k_i$
are defined through 
$m_Q/\bar m_Q(\bar{m}_Q) = 1 + k_1 \alpha_s/\pi+ k_2
\alpha_s^2/\pi^2 + k_3 \alpha_s^3/\pi^3 + \ldots$.
They can be found in Ref.~\cite{Gray:1990yh}
(see also
Refs.~\cite{Chetyrkin:1999qi,Chetyrkin:1999ys,Melnikov:2000qh,Marquard:2007uj}).

\begin{figure}[tbp]
\begin{center}
  \includegraphics[width=.48\textwidth]{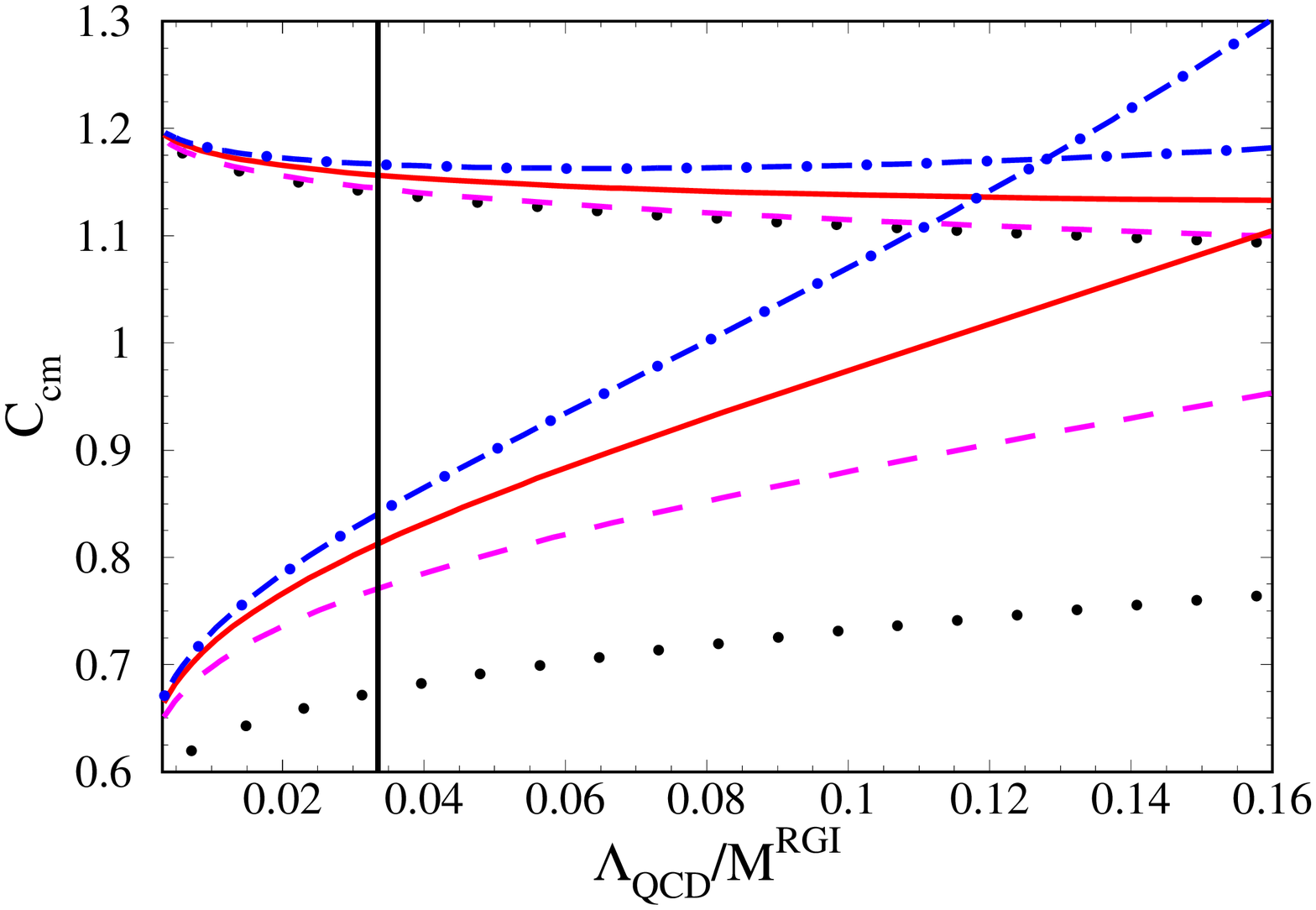}
  \includegraphics[width=.48\textwidth]{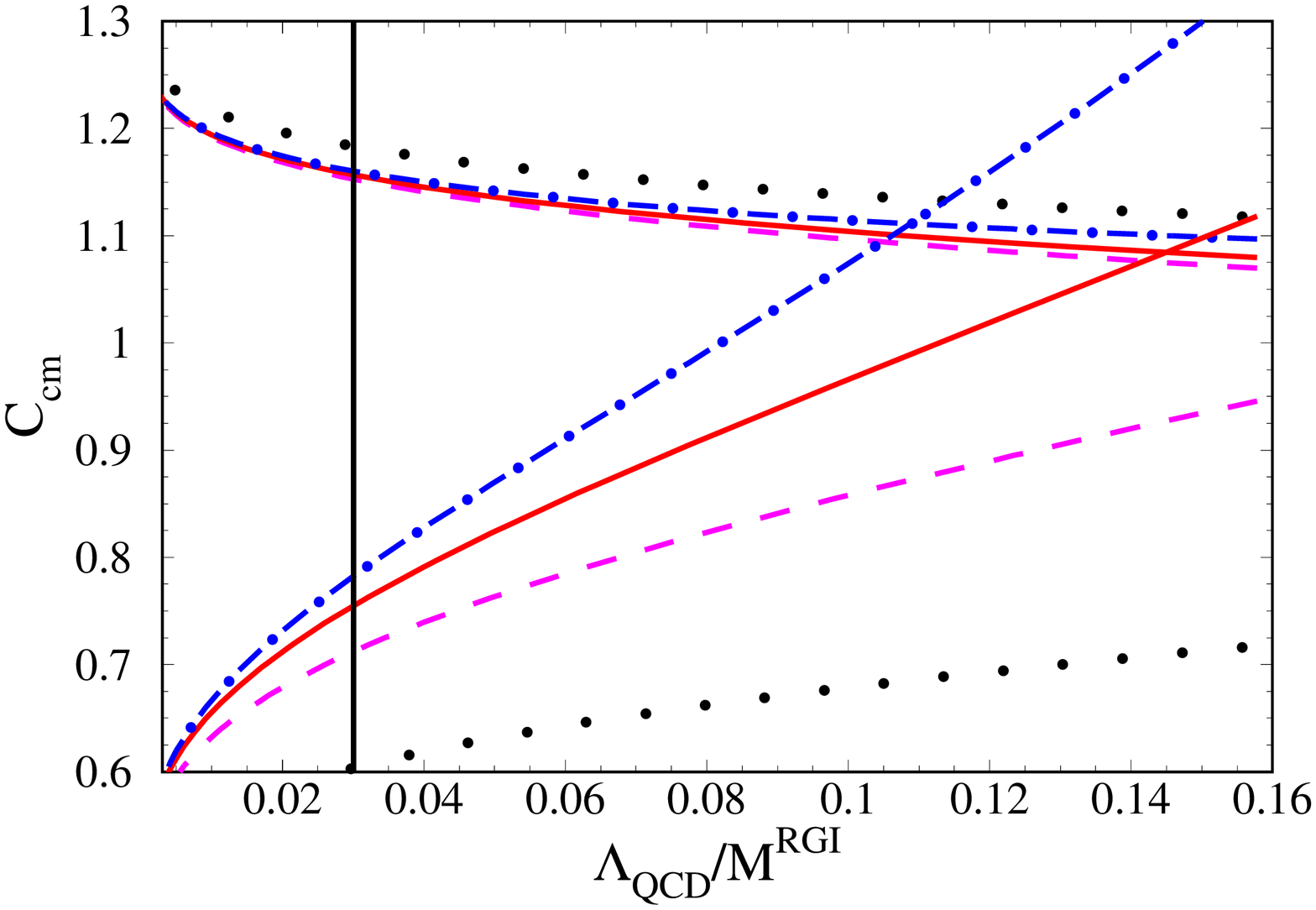}
\caption{\label{fig::cm}$C_{\rm mag}$ and $C_{\rm spin}$ as a function
  of $\Lambda_{\rm QCD}/m_Q$ for $n_l=0$ (left) and $n_l=4$ (right). The
  upper group of lines corresponds to $C_{\rm spin}$ while the lower
  shows $C_{\rm mag}$. Inside these groups the dotted, dashed
  and solid lines show the LO, NLO and NNLO, respectively. The
  estimation for the NNNLO result is shown by the dash-dotted line.
  The region relevant for the bottom quark is indicated by the
  vertical line.}
\end{center}
\end{figure}

In Fig.~\ref{fig::cm} the results for 
$C_{\rm mag}$ and $C_{\rm spin}$ are shown 
as a function of $\Lambda_{\rm QCD}/M^{RGI}$ for $n_l=0$ (left) and
$n_l=4$ (right) where the LO, NLO and NNLO results are shown. We also
added an estimation for the NNNLO result by assuming a vanishing
four-loop anomalous dimension which is motivated by the smallness of
the higher order terms in Eq.~(\ref{eq::gammacm-b}).
For the abscissa we choose $\Lambda_{\rm  QCD}=0.238$ which
results from a lattice calculation for $n_l=0$~\cite{Capitani:1998mq}
and vary the value for $M^{RGI}$.

For $\Lambda_{\rm QCD}/M^{RGI}\approx 0.03$, which is
the range relevant for the bottom quark, one observes in the case of
$C_{\rm mag}$ a relatively big shift when going from LO to NLO. However,
the additional shifts after including the NNLO and the (estimated)
NNNLO are smaller.
The convergence improves significantly when going from
$C_{\rm mag}$ to $C_{\rm spin}$. 
In the case of the bottom quark 
the NLO corrections turn out
to give a tiny contribution for $n_l=0$, however, 
also the NNLO and NNNLO results are quite small.
Going to smaller quark masses one observes moderate corrections for 
$C_{\rm spin}$ whereas for $C_{\rm mag}$ one has no convergence.
We would like to mention that in the case $n_l=4$ there is basically no
change in the behaviour of $C_{\rm mag}$. However, as far as 
$C_{\rm spin}$ is concerned one observes a moderate shift when
including the NLO terms whereas the NNLO and NNNLO corrections are tiny.

\section{Magnetic moment}
\label{sec::mag}

The calculation of the anomalous magnetic moment of a heavy quark 
proceeds along the same lines as for the chromomagnetic moment.
The only difference is that the external gluon is replaced by a
photon. Our result reads
\begin{eqnarray}
  \frac{a_Q}{Q_Q} &=& \frac{\alpha_s^{(n_f)}(m_Q)}{2 \pi} C_F \nonumber\\
  &&{} + \left( \frac{\alpha_s^{(n_f)}(m_Q)}{\pi} \right)^2 C_F \bigg[ \left(
  -\frac{31}{16} + \frac{5}{12} \pi^2 - \frac{1}{2} \pi^2 \ln 2 +
  \frac{3}{4} \zeta_3 \right) C_F \nonumber\\
  &&{} + \left(\frac{317}{144} - \frac{1}{8} \pi^2 + \frac{1}{4} \pi^2
  \ln 2 - \frac{3}{8} \zeta_3 \right) C_A \nonumber\\
  &&{} + \left( -\frac{25}{36} n_l
  + \frac{119}{36} - \frac{1}{3} \pi^2 \right) T_F \bigg]
  \nonumber\\
  &&{} + \left( \frac{\alpha_s^{(n_f)}(m_Q)}{\pi} \right)^3\, a_Q^{(3)} +
  \order{\alpha_s^4}\,.
  \label{eq::aq}
\end{eqnarray}
$Q_Q$ is the charge of the heavy quark
in terms of the positron charge.  Note that the strong coupling in
Eq.~(\ref{eq::aq}) is defined for $n_f$ active flavours and it is
evaluated at the scale $\mu=m_Q$. The two-loop contribution was
already computed in Ref.~\cite{Fleischer:1992re}. Recently, it has
also been obtained by considering the on-shell limit in the
calculation of the off-shell form factor~\cite{Bernreuther:2005gq}. We
are in full agreement with Ref.~\cite{Bernreuther:2005gq} while we
disagree with Ref.~\cite{Fleischer:1992re} by an overall factor of
four in the coefficient of $(\alpha_s/\pi)^2$.

Our new three-loop term is given by
\begin{eqnarray}
  a_Q^{(3)} &=& \bigg(
      \frac{20}{3} a_4
    + \frac{5}{18} \ln^4 2
    - \frac{5}{18} \pi^2 \ln^2 2
    - \frac{22}{3} \pi^2 \ln 2
    - \frac{235}{24} \zeta_5
    + \frac{103}{72} \pi^2 \zeta_3 \nonumber\\
  &&\qquad{}
    - \frac{139}{2160} \pi^4
    + \frac{241}{24} \zeta_3
    + \frac{23}{6} \pi^2
    - \frac{101}{64} 
  \bigg) C_F^3 \nonumber\\
  &&{} - \bigg(
      \frac{20}{3} a_4
    + \frac{5}{18} \ln^4 2
    - \frac{49}{36} \pi^2 \ln^2 2
    + \frac{31}{12} \pi^2 \ln 2
    - \frac{185}{24} \zeta_5
    + \frac{5}{12} \pi^2 \zeta_3 \nonumber\\
  &&\qquad{}
    + \frac{35}{432} \pi^4
    + \frac{113}{48} \zeta_3
    - \frac{1505}{432} \pi^2
    + \frac{955}{72}
  \bigg) C_F^2 C_A \nonumber\\
  &&{} + \bigg(
      \frac{5}{3} a_4
    + \frac{5}{72} \ln^4 2
    - \frac{11}{18} \pi^2 \ln^2 2
    + \frac{25}{8} \pi^2 \ln 2
    - \frac{65}{32} \zeta_5
    + \frac{29}{288} \pi^2 \zeta_3 \nonumber\\
  &&\qquad{}
    + \frac{103}{2880} \pi^4
    - \frac{1}{2} \zeta_3
    - \frac{463}{216} \pi^2
    + \frac{31231}{2592} 
  \bigg) C_F C_A^2 \nonumber\\
  &&{} - \bigg(
      \frac{8}{3} a_4
    + \frac{1}{9} \ln^4 2
    + \frac{2}{9} \pi^2 \ln^2 2
    - \frac{5}{3} \pi^2 \ln 2
    - \frac{11}{216} \pi^4
    + 3 \zeta_3 \nonumber\\
  &&\qquad{}
    + \frac{79}{54} \pi^2
    - \frac{125}{32}
  \bigg) C_F^2 T_F n_l \nonumber\\
  &&{} + \bigg(
      \frac{4}{3} a_4
    + \frac{1}{18} \ln^4 2
    + \frac{1}{9} \pi^2 \ln^2 2
    - \frac{5}{6} \pi^2 \ln 2
    - \frac{11}{432} \pi^4
    + \frac{19}{12} \zeta_3 \nonumber\\
  &&\qquad{}
    + \frac{77}{216} \pi^2
    - \frac{2411}{324}
  \bigg) C_F C_A T_F n_l \nonumber\\
  &&{} + \bigg(
      \frac{32}{3} a_4
    + \frac{4}{9} \ln^4 2
    - \frac{4}{9} \pi^2 \ln^2 2
    - \frac{16}{9} \pi^2 \ln 2
    + \frac{4}{135} \pi^4
    - \frac{263}{72} \zeta_3 \nonumber\\
  &&\qquad{}
    + \frac{11}{162} \pi^2
    + \frac{7703}{864}
  \bigg) C_F^2 T_F \nonumber\\
  &&{} - \bigg(
      \frac{20}{3} a_4
    + \frac{5}{18} \ln^4 2
    - \frac{5}{18} \pi^2 \ln^2 2
    - \frac{32}{9} \pi^2 \ln 2
    + \frac{25}{24} \zeta_5
    - \frac{1}{8} \pi^2 \zeta_3 \nonumber\\
  &&\qquad{}
    + \frac{143}{2160} \pi^4
    + \frac{241}{36} \zeta_3
    + \frac{1375}{648} \pi^2
    - \frac{2117}{162}
  \bigg) C_F C_A T_F \nonumber\\
  &&{} + \left[
      \frac{1}{27} \pi^2 (n_l + 1)
    + \frac{317}{324} n_l
    - \frac{61}{162}
  \right] C_F T_F^2 n_l \nonumber\\
  &&{} + \left(
      \frac{8}{3} \zeta_3
    - \frac{4}{135} \pi^2
    - \frac{943}{324}
  \right) C_F T_F^2
  + \frac{d_F^{abc}d_F^{abc}}{N_F}\, X_{\rm sin}\,,
\end{eqnarray}
where $d_F^{abc}$ is the symmetrized trace of three generators
in the fundamental representation (for SU$(N_c)$,
$d_F^{abc}d_F^{abc}=(N_c^2-1)(N_c^2-4)/(16N_c)$).
$X_{\rm sin}$ denotes the contribution from singlet diagrams
(cf. Fig.~\ref{fig::dias}(g)). It is given by
\begin{eqnarray}
  X_{\rm sin} &=&
      16 a_4
    + \frac{2}{3} \ln^4 2
    - \frac{2}{3} \pi^2 \ln^2 2
    - 24 \pi^2 \ln 2
    + \frac{5}{6} \zeta_5
    - \frac{5}{18} \pi^2 \zeta_3
    - \frac{41}{540} \pi^4 \nonumber\\
  &&\qquad{}
    - \frac{4}{3} \zeta_3
    + \frac{931}{54} \pi^2
    + \frac{5}{9} \,,
  \label{eq::light}
\end{eqnarray}
where we only include the contribution from diagrams with closed
heavy-quark loops. 
In principle there are contributions from diagrams
with massless quarks as well. However, within perturbation theory
they are divergent.  
Their evaluation either requires non-perturbative
methods or phenomenological models describing the interaction of 
light mesons in intermediate states.\footnote{In principle it is
  possible to calculate the contributions from massive charm and bottom
  quarks to $a_b$ and $a_t$, respectively. However, since the integrals
  involved contain two mass scales, this is beyond the scope of this
  work.}
This is in analogy to the
light-by-light contribution to the anomalous magnetic moment of the
muon~\cite{Laporta:1992pa}, which contains logarithms of the electron
mass.

It is possible to obtain the known three-loop result for the electron
anomalous magnetic moment from the expressions in
Eqs.~(\ref{eq::aq})--(\ref{eq::light}) by setting $C_F=T_F=1$, $C_A=0$,
$N_F=1$, $d_F^{abc} d_F^{abc}=1$ and $n_l=0$. This result was first obtained in
analytical form in Ref.~\cite{Laporta:1996mq} and confirmed in
Ref.~\cite{Melnikov:2000zc}.

Let us evaluate the quark magnetic moment numerically for charm, bottom
and top quarks. Inserting the numerical values for the coefficients, we
find
\begin{eqnarray}
  \frac{a_Q}{Q_Q} &=& 0.2122\, \alpha_s^{(n_f)}(m_Q) + \left( 0.8417 -
  0.0469\, n_l \right) \left[\alpha_s^{(n_f)}(m_Q)\right]^2 \nonumber\\
  &&{} + \left( 4.5763 - 0.5856\, n_l + 0.0145\, n_l^2 \right)
  \left[\alpha_s^{(n_f)}(m_Q)\right]^3 + \order{\alpha_s^4} \,.
\end{eqnarray}
Using 
$\alpha_s^{(4)}(m_c)=0.3378$, $\alpha_s^{(5)}(m_b)=0.2169$ and
$\alpha_s^{(6)}(m_t)=0.1075$ we obtain
\begin{eqnarray}
  a_c &=& \hphantom{-}0.0478 + 0.0533 + 0.0758 = \hphantom{-}0.1770\,, 
  \nonumber\\
  a_b &=& -0.0153 - 0.0103 - 0.0084 = -0.0340\,, 
  \nonumber\\
  a_t &=& \hphantom{-}0.0152 + 0.0047 + 0.0017 = \hphantom{-}0.0215\,.
\end{eqnarray}
The pattern is very similar to the chromomagnetic moment of the 
heavy quark: no convergence is observed for the charm quark,
the corrections for the bottom quark are large and amount to
more than 50\% of the one-loop contribution, and
in the top quark case, we find that our new term 
gives a 10\% contribution which provides quite some confidence
that the uncertainty of the final prediction for $a_t$ is small.

As already mentioned in the Introduction 
the LEP1 bound of Ref.~\cite{Escribano:1993xr}
for the bottom quark --- $a_b/Q_b<0.045$ (68\%C.L. ) --- was
found to be saturated by the two-loop correction. It is therefore
interesting to see what happens if we include our three-loop term. For
this purpose, we have to evaluate $a_b$ for $\mu=m_Z$. We find
\begin{equation}
  a_b(m_Z) = -0.0083 - 0.0066 - 0.0056 = -0.0206\,.
\end{equation}
Since the three-loop correction is almost as large as the two-loop one
this overshoots the bound by about 25\%. In this context we want to
mention again 
that the contributions from closed light-quark loops could not be included
in our calculation. These contributions might decrease the three-loop
correction. In any case, a more precise measurement of $a_b$ would
certainly be interesting.


\section{Conclusion}
\label{sec::conc}

Our main results are the three-loop anomalous dimension
of the HQET chromo\-magnetic-interaction operator~(\ref{eq::gamma})
and the three-loop coefficient of this operator
in the HQET Lagrangian~(\ref{eq::Cm}).
They can be used for the investigation of various effects
of the heavy-quark spin symmetry violation
(e.g., mass splittings) using continuum or lattice techniques.
Furthermore, we have obtained the three-loop anomalous magnetic moments
of heavy quarks~(\ref{eq::aq}).
This result does not include the light-quark light-by-light contribution
which cannot be calculated perturbatively.


\section*{Acknowledgements}

We would like to thank Rainer Sommer for fruitful discussions, many
explanations in connection to $C_{\rm mag}$ and $C_{\rm spin}$ and
carefully reading the manuscript.
We also thank Luminita Mihaila for discussions about the colour factors.
The work of J.P. was partially supported by NSERC.
This work was supported by the DFG through SFB/TR~9.
The Feynman diagrams were drawn with {\tt JaxoDraw}~\cite{Binosi:2003yf}.




\begin{thebibliography}{99}

%
%

\bibitem{Eichten:1990vp}
  E.~Eichten and B.~R.~Hill,
  Phys.\ Lett.\  B {\bf 243} (1990) 427.

\bibitem{Falk:1990pz}
  A.~F.~Falk, B.~Grinstein and M.~E.~Luke,
  Nucl.\ Phys.\  B {\bf 357} (1991) 185.

\bibitem{Manohar:2000dt}
  A.~V.~Manohar and M.~B.~Wise,
  \textit{Heavy Quark Physics},
  Camb.\ Monogr.\ Part.\ Phys.\ Nucl.\ Phys.\ Cosmol.\  {\bf 10} (2000) 1.

\bibitem{Grozin:2004yc}
  A.~G.~Grozin,
  \textit{Heavy Quark Effective Theory},
  Springer Tracts Mod.\ Phys.\  {\bf 201} (2004) 1.

\bibitem{Luke:1992cs}
  M.~E.~Luke and A.~V.~Manohar,
  Phys.\ Lett.\  B {\bf 286} (1992) 348
  [arXiv:hep-ph/9205228].

\bibitem{Abbott:1980hw}
  L.~F.~Abbott,
  Nucl.\ Phys.\  B {\bf 185} (1981) 189.

\bibitem{Amoros:1997rx}
  G.~Amor\'os, M.~Beneke and M.~Neubert,
  Phys.\ Lett.\  B {\bf 401} (1997) 81
  [arXiv:hep-ph/9701375].

\bibitem{Czarnecki:1997dz}
  A.~Czarnecki and A.~G.~Grozin,
  Phys.\ Lett.\  B {\bf 405} (1997) 142
  [arXiv:hep-ph/9701415].

\bibitem{Grozin:1997ih}
  A.~G.~Grozin and M.~Neubert,
  Nucl.\ Phys.\  B {\bf 508} (1997) 311
  [arXiv:hep-ph/9707318].

\bibitem{Davydychev:1998si}
  A.~I.~Davydychev and A.~G.~Grozin,
  Phys.\ Rev.\  D {\bf 59} (1999) 054023
  [arXiv:hep-ph/9809589].

\bibitem{Sommer:2006sj}
  R.~Sommer,
  arXiv:hep-lat/0611020.

\bibitem{Guazzini:2007bu}
  D.~Guazzini, H.~B.~Meyer and R.~Sommer  [ALPHA Collaboration],
  arXiv:0705.1809 [hep-lat].

\bibitem{Escribano:1993xr}
  R.~Escribano and E.~Mass\'{o},
  Nucl.\ Phys.\  B {\bf 429} (1994) 19
  [arXiv:hep-ph/9403304].

\bibitem{Bernreuther:2005gq}
  W.~Bernreuther, R.~Bonciani, T.~Gehrmann, R.~Heinesch, T.~Leineweber, P.~Mastrolia and E.~Remiddi,
  Phys.\ Rev.\ Lett.\  {\bf 95} (2005) 261802
  [arXiv:hep-ph/0509341].

\bibitem{Melnikov:2000zc}
  K.~Melnikov and T.~van Ritbergen,
  Nucl.\ Phys.\  B {\bf 591} (2000) 515
  [arXiv:hep-ph/0005131].

\bibitem{Marquard:2007uj}
  P.~Marquard, L.~Mihaila, J.~H.~Piclum and M.~Steinhauser,
  Nucl.\ Phys.\  B {\bf 773} (2007) 1
  [arXiv:hep-ph/0702185].

\bibitem{Nogueira:1991ex}
  P.~Nogueira,
  J.\ Comput.\ Phys.\  {\bf 105} (1993) 279.

\bibitem{Harlander:1997zb}
  R.~Harlander, T.~Seidensticker and M.~Steinhauser,
  Phys.\ Lett.\  B {\bf 426} (1998) 125
  [arXiv:hep-ph/9712228].

\bibitem{Seidensticker:1999bb}
  T.~Seidensticker,
  arXiv:hep-ph/9905298.

\bibitem{Laporta:1996mq}
  S.~Laporta and E.~Remiddi,
  Phys.\ Lett.\  B {\bf 379} (1996) 283
  [arXiv:hep-ph/9602417].

\bibitem{Laporta:2001dd}
  S.~Laporta,
  Int.\ J.\ Mod.\ Phys.\  A {\bf 15} (2000) 5087
  [arXiv:hep-ph/0102033].

\bibitem{PMDS}
  P.~Marquard and D.~Seidel,
  unpublished.

\bibitem{Bauer:2000cp}
  C.~Bauer, A.~Frink and R.~Kreckel,
  arXiv:cs.sc/0004015.

\bibitem{fermat} R.~H.~Lewis, Fermat's User Guide,
  http://www.bway.net/\~{}lewis.

\bibitem{Tentyukov:2006ys}
  M.~Tentyukov and J.~A.~M.~Vermaseren,
  Comput.\ Phys.\ Commun.\  {\bf 176} (2007) 385
  [arXiv:cs.sc/0604052].

\bibitem{Chetyrkin:1981qh}
  K.~G.~Chetyrkin and F.~V.~Tkachov,
  Nucl.\ Phys.\  B {\bf 192} (1981) 159.

\bibitem{vanRitbergen:1998pn}
  T.~van Ritbergen, A.~N.~Schellekens and J.~A.~M.~Vermaseren,
  Int.\ J.\ Mod.\ Phys.\  A {\bf 14} (1999) 41
  [arXiv:hep-ph/9802376].

\bibitem{Chetyrkin:1997un}
  K.~G.~Chetyrkin, B.~A.~Kniehl and M.~Steinhauser,
  Nucl.\ Phys.\  B {\bf 510} (1998) 61
  [arXiv:hep-ph/9708255].

\bibitem{Chetyrkin:2000yt}
  K.~G.~Chetyrkin, J.~H.~K{\"u}hn and M.~Steinhauser,
  Comput.\ Phys.\ Commun.\  {\bf 133} (2000) 43
  [arXiv:hep-ph/0004189].

\bibitem{Bigi:1994ga}
  I.~I.~Y.~Bigi, M.~A.~Shifman, N.~G.~Uraltsev and A.~I.~Vainshtein,
  Phys.\ Rev.\  D {\bf 52} (1995) 196
  [arXiv:hep-ph/9405410].

\bibitem{Grozin:2000cm}
  A.~G.~Grozin,
  arXiv:hep-ph/0008300.

\bibitem{Grozin:2006xm}
  A.~G.~Grozin, A.~V.~Smirnov and V.~A.~Smirnov,
  JHEP {\bf 0611} (2006) 022
  [arXiv:hep-ph/0609280].

\bibitem{Grozin:2007ap}
  A.~G.~Grozin, T.~Huber and D.~Maitre,
  arXiv:0705.2609 [hep-ph].

\bibitem{vanRitbergen:1997va}
  T.~van Ritbergen, J.~A.~M.~Vermaseren and S.~A.~Larin,
  Phys.\ Lett.\  B {\bf 400} (1997) 379
  [arXiv:hep-ph/9701390].

\bibitem{Czakon:2004bu}
  M.~Czakon,
  Nucl.\ Phys.\  B {\bf 710} (2005) 485
  [arXiv:hep-ph/0411261].

\bibitem{pdg}
  W.~M.~Yao {\it et al.} [Particle Data Group],
  J.\ Phys.\ C {\bf 33} (2006) 1,\\
  URL: http://pdg.lbl.gov.

\bibitem{Heitger:2004gb}
  J.~Heitger, A.~Juttner, R.~Sommer and J.~Wennekers  [ALPHA Collaboration],
  JHEP {\bf 0411} (2004) 048
  [arXiv:hep-ph/0407227].

\bibitem{Chetyrkin:1997dh}
  K.~G.~Chetyrkin,
  Phys.\ Lett.\  B {\bf 404} (1997) 161
  [arXiv:hep-ph/9703278].

\bibitem{Vermaseren:1997fq}
  J.~A.~M.~Vermaseren, S.~A.~Larin and T.~van Ritbergen,
  Phys.\ Lett.\  B {\bf 405} (1997) 327
  [arXiv:hep-ph/9703284].

\bibitem{Gray:1990yh}
  N.~Gray, D.~J.~Broadhurst, W.~Grafe and K.~Schilcher,
  Z.\ Phys.\  C {\bf 48} (1990) 673.

\bibitem{Chetyrkin:1999qi}
  K.~G.~Chetyrkin and M.~Steinhauser,
  Nucl.\ Phys.\  B {\bf 573} (2000) 617
  [arXiv:hep-ph/9911434].

\bibitem{Chetyrkin:1999ys}
  K.~G.~Chetyrkin and M.~Steinhauser,
  Phys.\ Rev.\ Lett.\  {\bf 83} (1999) 4001
  [arXiv:hep-ph/9907509].

\bibitem{Melnikov:2000qh}
  K.~Melnikov and T.~v.~Ritbergen,
  Phys.\ Lett.\  B {\bf 482} (2000) 99
  [arXiv:hep-ph/9912391].

\bibitem{Capitani:1998mq}
  S.~Capitani, M.~Luscher, R.~Sommer and H.~Wittig  [ALPHA
  Collaboration],
  Nucl.\ Phys.\  B {\bf 544} (1999) 669
  [arXiv:hep-lat/9810063].

\bibitem{Fleischer:1992re}
  J.~Fleischer and O.~V.~Tarasov,
  Phys.\ Lett.\  B {\bf 283} (1992) 129.

\bibitem{Laporta:1992pa}
  S.~Laporta and E.~Remiddi,
  Phys.\ Lett.\  B {\bf 301} (1993) 440.

\bibitem{Binosi:2003yf}
  D.~Binosi and L.~Theussl,
  Comput.\ Phys.\ Commun.\  {\bf 161} (2004) 76
  [arXiv:hep-ph/0309015].


\end{thebibliography}
\end{document}